\documentclass[12pt]{amsart}
\usepackage{amscd}
\usepackage{mathrsfs}
\newtheorem{theorem}{Theorem}[section]

\newtheorem{corollary}[theorem]{Corollary}
\theoremstyle{definition}

\newtheorem{remark}[theorem]{Remark}
\numberwithin{equation}{section}

\def\cd{\!\cdot \!}
\def\D{\mathscr D}

\def\RE{\mathbb R}
\def\CO{{\mathbb C}}
\def\S{\mathscr S}

\def\C{\mathscr C}
\def\L{\mathscr L}

\def\min{\text{\rm min}}
\def\max{\text{\rm max}}
\newsymbol\subsetneq 2328

\begin{document}

\title[Hamiltonian operators for
  systems with boundaries]
{Self-adjoint, globally defined Hamiltonian operators for
  systems with boundaries }

\author{Nuno Costa Dias}
\address{Universidade Lus\'ofona de Humanidades e
Tecnologias, Av. Campo Grande 376, 1749-024 Lisboa,
Portugal and Grupo de F\'{\i}sica Matem\'atica,
Universidade de Lisboa, Av. Prof. Gama Pinto 2, 1649-003,
Lisboa, Portugal}
\email{ncdias@meo.pt, joao.prata@ulusofona.pt}

\author{Andrea Posilicano}
\address{Dipartimento di Scienze Fisiche e Matematiche,  Universit\`a
dell'Insubria, I-22100 Como, Italy}
\email{posilicano@uninsubria.it}

\author{Jo\~{a}o Nuno Prata}

%\keywords{Self-Adjoint Extensions, Kre\u \i n's Resolvent Formula,
%Elliptic Differential Operators}
%\thanks{{\it Mathematics Subject Classification (2000).} 47B25
%(primary),
%47B38, 35J25 (secondary)}

\begin{abstract}
For a general self-adjoint Hamiltonian operator $H_0$ on the
Hilbert space $L^2(\RE^d)$, we determine the set of all
self-adjoint Hamiltonians $H$ on $L^2(\RE^d)$ that dynamically
confine the system to an open set $\Omega \subset \RE^d$ while
reproducing the action of $ H_0$ on an appropriate operator
domain. In the case $H_0=-\Delta +V$ we construct these
Hamiltonians explicitly showing that they can be written in the
form $H=H_0+ B$, where $B$ is a singular boundary potential and
$H$ is self-adjoint on its maximal domain. An application to the
deformation quantization of one-dimensional systems with
boundaries is also presented.

\end{abstract}

{\maketitle }
\begin{section}{Introduction.}
This paper concerns the quantum formulation of systems with
boundaries. These systems play an important part in several fields
of current research like, for instance, in mathematical physics
(e.g. the theory of self-adjoint extensions of symmetric operators
\cite{Neumann, [K1],[K2], Faris,Reed,Akhiezer,Naimark,[BS],[Weid],
Posilicano1,Bonneau,Voronov,Garbaczewski,Posilicano2,[BEH],
[deO]}), condensed matter physics (e.g. the quantum description of
particles moving on surfaces with obstacles or impurities
\cite{Kowalski,Cacciapuoti}) and in string theory
\cite{Witten,Pinzul} and other modern approaches to quantum
gravity \cite{Isham,Gambinni} (where the classical theory displays
a non-trivial global structure \cite{Isham}).
\par
Let us consider a $d$-dimensional dynamical system confined to an
open set $\Omega \subset \RE^d$. Two main approaches to the
canonical quantization of these systems \cite{Garbaczewski}, are:
\\

(A) The kinematical approach, where the confinement is a
consequence of the choice of the Hilbert space, assumed to be
$L^2(\Omega)$.
\\

(B) The dynamical approach, where the system is formulated in the
unconfined Hilbert space $L^2(\RE^d)$ and the confinement is a
feature of dynamics i.e. it is a consequence of the choice of the
Hamiltonian operator.
\\

In both cases one is faced with the problem of determining
self-adjoint (s.a.) realizations of the Hamiltonian operator (i.e.
to determine a formal s.a. differential expression and a domain
such that $H=H^*$). Notice that the implementation of the
Hamiltonian operator (as well as other fundamental observables)
only as a symmetric operator ($S\subset S^*$) does not yield a
well defined physical observable (see e.g. \cite{[vN]}). The
difference between symmetric and s.a. operators is an important
one but also a subtle one. In the context of the approach (A) this
difference can only be realized from a careful analysis of the
operator domains. On the other hand, in the context of (B), the
situation is more transparent as the symmetric Hamiltonian, its
adjoint and each of its s.a. realizations are naturally defined by
different differential expressions.\par

A more detailed analysis also shows that, at a fundamental level,
the (more standard) approach (A) reveals unexpected
inconsistencies \cite{Garbaczewski,Bonneau,Isham}. These are
mainly related to ambiguities in the physical predictions (when
there are several possible self-adjoint realizations of a single
observable), to the absence of self-adjoint (s.a.) formulations of
important observables and to difficulties in translating this
approach to other (non-local) formulations of quantum mechanics,
like the deformation or the de Broglie-Bohm formulations
\cite{Dias1,Kryukov,Dias6,Walton}.

These problems are well illustrated by the textbook example of a
one-dimensional single particle with Hamiltonian $
H=-\frac{d^2}{dx^2}$, and confined to a half-line or to a finite
interval \cite{Akhiezer,Garbaczewski,Bonneau} (and they are also
present, in alternate forms, in the higher dimensional case).
Taking the example of the positive half-line case, they can be
summarized as follows: i) There is no s.a. operator acting as
$-i\frac{d}{dx}$ on a dense subspace of $L^2(\RE^+)$ and thus the
momentum is not an observable of this system
\cite{Garbaczewski,Bonneau}. ii) Since there is no momentum, the
Hamiltonian cannot be defined by $H=\frac{p^2}{2m}$. iii) On the
other hand, the direct definition $H=-\frac{d^2}{dx^2}$ is also
ambiguous because there is an all one-parameter family of s.a.
realizations of the differential expression $-\frac{d^2}{dx^2}$ on
$L^2(\RE^+)$, each of which determines a particular dynamics and a
different set of physical properties (see \cite{Bonneau}, section 6).
The complete specification of $H$ requires a particular choice of
boundary conditions at $x=0$, but it is difficult to see what is
the physics behind a choice of the boundary condition. iv) Some of
the allowed s.a. boundary conditions lead to formulations
displaying some surprising properties, like the existence of
negative energy eigenstates, in spite of the fact that the
Hamiltonian is formally a positive defined operator
\cite{Bonneau}. v) Finally, the deformation quantization (as other
non-local formulations of quantum mechanics) of the "kinematical"
formulation of the system is problematic
\cite{Dias1,Kryukov,Dias6,Walton}. In section IV we will discuss
this last point in detail.

The approach (B), on the other hand, displays the obvious
advantage that the most important observables (like the momentum)
are naturally defined as s.a. operators and further, that there is
no ambiguity on the choice of the s.a. boundary conditions, as
they are completely determined by the particular boundary
potential displayed by each s.a. Hamiltonian. Finally, as we will
see in section 4, the approach (B) is crucial to address the
deformation quantization of systems with boundaries. The main
problem in (B) is the explicit construction of the operators,
defined on $L^2(\RE^d)$ but effectively confining the system to
its domain $\Omega \subset \RE^d$. This approach has been scarcely
explored in the literature. Up to our knowledge, one of the few
references in the subject is \cite{Garbaczewski} where the authors
propose and study some of the features of a mechanism for
dynamical confinement. Some related work on the relations between
partially and globally defined operators was presented in
\cite{Voronov}.

In this paper we shall further study the dynamical confinement
point of view. The problems that will be addressed are closely
related to the topics of singular perturbations of s.a. operators
\cite{Albeverio,Posilicano1}, point interaction Hamiltonians
\cite{Albeverio,Berezin,Ph} and s.a. extensions of symmetric
restrictions \cite{Garbaczewski,Posilicano2}. Our starting point
will be a generic unconfined dynamical system defined on the
Hilbert space $L^2(\RE^d)$ and described by a s.a. Hamiltonian $
H_0$. \par Given an open set $\Omega\subset \RE^d$ and denoting by
$\chi_\Omega$ its characteristic function, we consider the
orthogonal projection
$$
P_\Omega:L^2(\RE^d)\to L^2(\RE^d)\,,\quad P_\Omega \psi=
\chi_\Omega \psi \,,
$$
so that
$$
L^2(\RE^d)\simeq{\rm Ran}(P_\Omega)\oplus{\rm Ker}(P_\Omega)
\equiv L^2(\Omega)\oplus L^2(\Omega^c)\,.
$$
This paper is devoted to solving the two following problems:
\vskip 8pt\noindent {\bf Problem 1.} {\it Given a s.a. linear
operator $$ H_0:D(H_0)
  \subseteq L^2(\RE^d) \to L^2(\RE^d)\,,$$ determine the explicit form
  of all linear operators $$H:D(H) \subseteq L^2(\RE^d) \to
  L^2(\RE^d)$$ that satisfy the following three properties:
\par\noindent
(1) $P_\Omega (D(H))\subseteq D(H)$ and $[ P_\Omega, H]\psi=0$ for
all $\psi\in D(H)$;
\par\noindent
(2) $H$ is self-adjoint;
\par\noindent
(3) if $\psi \in D(H_0)$ is an eigenstate of $ P_\Omega$ then
$\psi\in D(H)$ and $ H\psi = H_0\psi$. } \vskip 8pt\noindent We
will refer to the operators $H$ as the {\it confining
Hamiltonians} and to the properties (1) to (3) as the {\it
defining properties of $ H$}.
\par
From (1) and (2) we find that $ P_\Omega$ commutes with all the
spectral projectors of $ H$ and so also with the operator $e^{ -i
t H  }$. Hence, if $\psi $ is an eigenstate of $ P_\Omega$ (with
eigenvalue $0$ or $1$) it will evolve to $e^{-it  H  }\psi $,
which is again an eigenstate of $ P_\Omega$ with the same
eigenvalue. In other words, $ P_\Omega$ is a constant of motion
and a wave function confined to $\Omega$ (or to $\Omega^c$) will
stay so forever. Finally, property (3) imposes that, for wave
functions in $D(H_0)$ with support on a subset of $\Omega$ (or
$\Omega^c$), i.e. for wave functions that do not "see" the
boundary, the infinitesimal time evolution determined by $H$
reproduces the original one given by $H_0$. \vskip 8pt\noindent
{\bf Problem 2.} {\it For $H_0= -\Delta +V$ determine whether it
is possible to write the corresponding confining Hamiltonians $ H$
(solutions of Problem 1) in the form $ H= H_0+ B$ where $ B$ is a
singular boundary potential and $ H$ is s.a. on its maximal
domain.} \vskip 8pt\noindent The first part of this paper (section
2) is devoted to Problem 1. We shall characterize the operators
that satisfy properties (1) to (3), determine the properties that
the original $ H_0$ should satisfy so that the operators $ H$ do
exist and derive a method to construct these operators explicitly.
In this context we will also explore the relations between
partially and globally defined operators. The results of this
section lead naturally to the construction of yet another class of
s.a. Hamiltonians, which describe systems composed of separate
domains but allow for some sort of information transfer between
these domains. Further investigation on these operators will be
left for a future work \cite{Dias4}.

The second part of the paper (section 3) is devoted to Problem 2.
We specialize to Hamiltonians of the form $H_0=-\Delta +V$ and
introduce a new kind of singular operators. These will be used to
write the Hamiltonians $H$, satisfying the three defining
properties, in the desired form: $ H= H_0 + B$, where $ B$ is a
singular boundary potential which is dependent of the boundary
conditions that characterize the domain of $ H$. This is always
possible. Indeed $H$ amounts to a self-adjoint extension of the
symmetric restriction of $H_0$ to the domain
$D(\Delta_{\Omega_1}^{\min})\oplus D(\Delta_{\Omega_2}^{\min})$,
where $\Delta_{\Omega_k}^{\min}$, $k=1,2$, denotes the minimal
Laplacian on $\Omega_k$ with operator domain given by the set of
smooth functions with compact support contained in $\Omega_k$,
$\Omega_1=\Omega$, $\Omega_2=(\bar\Omega)^c$. Thus, by the
additive representation of self-adjoint extensions obtained in
\cite{[Po]} (also see \cite{Posilicano2}), $H$ admits the additive
representation $H=-\Delta+V+B$, where $B$ is a singular boundary
potential which we explicitly determine. Such a singular boundary
potential is defined in terms of the zero'th and first order trace
operators on the boundary or better of their extensions (provided
in \cite{[LM-AIF],[LM]}) to the maximal domains
$D(\Delta_{\Omega_k}^{\max})=\{\psi_k\in
  L^2(\Omega_k)\,:\,\Delta_{\Omega_k}\psi_k\in L^2(\Omega_k)\}$.
\par
As a final result of section 3, we show that the operators $ H$
are s.a. on their maximal domain. Hence, contrary to what is
common in the approach (A), there is no ambiguity regarding the
boundary conditions satisfied by the domain of $ H$. In fact, each
$ H$ satisfying (1) to (3) exhibits a particular functional form
$H= H_0 + B$ (it displays a particular boundary potential $B$) and
its self-adjointness domain turns out to be  its maximal domain.

Finally, in section 4 we provide a simple example in order to
illustrate the general results of the previous sections. We
determine the explicit form of the boundary potential for a
one-dimensional free particle, confined to an interval and subject
to Dirichlet boundary conditions. We also discuss the problems
involved in the deformation formulation of confined systems and
use our simple example to show that, in the deformation context, a
consistent formulation of the (confined) energy eigenvalue problem
can only be obtained using the boundary potential approach.

Lastly, let us point out that there are some interesting topics
related to the results of this paper that could be studied. These
may include: the global formulation of systems composed by several
domains and displaying some kind of information transfer between
different domains; the application of the results of this paper to
the deformation quantization of higher dimension confined systems
\cite{Bayen,Kontsevich,Dias1,Kryukov} and to the noncommutative
formulation of manifolds with boundaries \cite{Connes,Madore}.

\end{section}

\begin{section}{Confining Hamiltonians on $L^2(\RE^d)$}

In this section we will study the operators $$ H:D(H) \subseteq
L^2(\RE^d) \to L^2(\RE^d)$$ associated to a s.a. $H_0$, and satisfying
the defining properties (1) to (3). We will prove that all these operators are of the form $ H= H_1 \oplus  H_2$ where $H_1$ and $ H_2$ are s.a
extensions of the restrictions of $ H_0$ to a suitable domain.
\par
Given the open set $\Omega\subset \RE^d$, we pose
$$\Omega_1:=\Omega\,,\qquad \Omega_2:=(\bar\Omega_1)^c$$
and use the decomposition of the orthogonal projection
$$
P_k\equiv P_{\Omega_k}=E_kR_k\,,\quad E_k=R_k^*\,,\quad k=1,2\,,$$
given by the
restriction and extension operators
$$
R_k:L^2(\RE^d)\to L^2(\Omega_k)\,,\quad [R_k\psi](x):=\psi(x)
\,,\ x\in\Omega_k\,,
$$
$$
E_k:L^2(\Omega_k)\to L^2(\RE^d)\,,\quad [E_k\psi_k](x):=
\begin{cases}\psi_k(x) & x\in \Omega_k\\
0 & x\in \Omega_k^c\,.
\end{cases}
$$
By such operators one has the identification
$$L^2(\Omega_1)\oplus L^2(\Omega_2)\simeq L^2(\RE^d)$$ given
by the unitary map
$$
J:L^2(\Omega_1)\oplus L^2(\Omega_2)\to L^2(\RE^d)\,,
\quad J(\psi_1\oplus\psi_2):=E_1\psi_1+E_2\psi_2\,,
$$
with inverse
$$
J^{-1}:L^2(\RE^d)\to L^2(\Omega_1)\oplus L^2(\Omega_2)\,,
\quad J^{-1}\psi:=R_1\psi\oplus R_2\psi\,.
$$
Given two linear operators
$$L_k:D(L_k)\subseteq L^2(\Omega_k)\to L^2(\Omega_k), ~ k =1,2$$ we pose
as usual
$$
L_1\oplus L_2:D(L_1)\oplus D(L_2)\subseteq
L^2(\Omega_1)\oplus L^2(\Omega_2)\to
L^2(\Omega_1)\oplus L^2(\Omega_2)\,,
$$
$$
L_1\oplus L_2\,\psi_1\oplus\psi_2:=L_1\psi_1\oplus L_2\psi_2\,.
$$
Given any two subspaces $V_k\subseteq L^2(\Omega_k)$,
we define the subspace $V_1\tilde\oplus V_2\subseteq L^2(\RE^d)$ by
$$
V_1\tilde\oplus V_2:=J(V_1\oplus V_2)\,,
$$
and then the operator on $L^2(\RE^d)$
$$
L_1\tilde\oplus L_2 :D(L_1)\tilde\oplus D(L_2)\subseteq L^2(\RE^d)\to L^2(\RE^d)\,,
$$
$$
L_1\tilde\oplus L_2:= J(L_1\oplus L_2)J^{-1}\,.
$$
Then one has the following
\begin{theorem} A linear operator $ H:D(H) \subseteq L^2(\RE^d) \to L^2(\RE^d)$ satisfies the defining property (1) above iff it can be written in the form
\begin{equation*}
 H= H_1\tilde \oplus  H_2 : D(H_1)\tilde \oplus D(H_2)
\subseteq L^2(\RE^d)\to L^2(\RE^d)\,,
\end{equation*}
where
$$
H_k:D(H_k):= R_k(D(H))\subseteq L^2(\Omega_k)\to L^2(\Omega_k)\,,\quad
H_k:=R_kHE_k \,.
$$
\end{theorem}

\begin{proof} By known results on reducing subspaces
(see e.g. \cite{[BS]}, Chapter 3, Section 6) one has that property
(1) holds true iff $L^2(\Omega)$ reduces $H$. Thus
$$
HP_k(D(H))\subseteq P_k(L^2(\RE^d))
$$
and
$$
H\psi=P_1HP_1\psi+P_2HP_2\psi\,.
$$
Hence
\begin{align*}
H\psi=&E_1R_1HE_1R_1\psi+E_2R_2HE_2R_2\psi\\
=&
J(R_1HE_1R_1\psi+R_2HE_2R_2\psi)\\
=&J(R_1HE_1\oplus R_2HE_2)J^{-1}\psi\,.
\end{align*}
\end{proof}

\begin{remark} By the known properties of direct sums of operators,
  and since $H_1\oplus H_2$ and $H_1\tilde\oplus H_2$ are
unitarily
  equivalent, one has that
$$
\text{\rm $ H$ is symmetric iff $ H_1$ and $ H_2$
are symmetric,}
$$
and, in the case both $D(H_1)$ and $D(H_2)$ are dense,
 $$
H^*=H_1^* \tilde\oplus  H_2^*\,.$$
Be aware that here and below by symmetric operator we just mean a
linear operator $S$ such that
$\langle S\phi,\psi\rangle=\langle\phi,S\psi\rangle$ for all
$\phi,\psi\in D(S)$; the operator domain
$D(S)$ could be not dense.   \par
In particular $ H$ is self-adjoint iff $ H_1 $ and $ H_2$ are both
self-adjoint. Thus a Hamiltonian $ H$ satisfies the defining
properties (1) and (2) iff it can be written in the form $ H=
H_1\tilde\oplus  H_2$ where $ H_1$ and $ H_2$ are given in the
previous theorem and self-adjoint.
\end{remark}

We now investigate the implications of the defining
property (3).\par
Let us consider the operators
$$
S_k:D(S_k)\subseteq L^2(\Omega_k)\to
L^2(\Omega_k)\,,\quad S_k:=R_kH_0E_k
$$
$$
D(S_k):=\{\psi_k\in L^2(\Omega_k)\,:\, E_k\psi_k\in
D(H_0)\}\,.
$$
For any $\phi_k,\psi_k\in D(S_k)$ one has
\begin{align*}
&\langle R_kH_0E_k\phi_k,\psi_k\rangle_{L^2(\Omega_k)}=
\langle H_0E_k\phi_k,E_k\psi_k\rangle_{L^2(\RE^d)}\\
=&\langle E_k\phi_k,H_0E_k\psi_k\rangle_{L^2(\RE^d)}=
\langle \phi_k,R_kH_0E_k\psi_k\rangle_{L^2(\Omega_k)}
\end{align*}
and so
$S_k$ is a symmetric
operator.
Then is immediate to check
that property (3) is equivalent to
$$
\tilde H_0\subseteq H\,,
$$
where $\tilde H_0$ denotes the symmetric operator given by the
restriction of $H_0$ to $D(S_1)\tilde\oplus D(S_2)$.
Thus (1) puts the constraint
\begin{equation}\label{commutator}
[P_\Omega,\tilde H_0]=0
\end{equation}
on $H_0$. By (\ref{commutator}) and Theorem 2.1 one gets
$$
\tilde H_0=S_1\tilde\oplus S_2
$$
and so
\begin{theorem} Let $H_0$ be a s.a. operator. If $[P_\Omega,\tilde H_0]\not=0$ then there is no operator $H$ satisfying (1) to (3). If, on the other hand,
$[P_\Omega,\tilde H_0]=0$ then $H$ satisfies (1) to (3) if and only if the symmetric operators
$S_k$ admit self-adjoint extensions $H_k$ and $H=H_1\tilde \oplus H_2$.
\end{theorem}

\begin{remark}
Let us suppose that both the domains of the symmetric operators
$S_1$ and $S_2$ are dense. From Remark 2.2 we know that the adjoint of
$S_1 \tilde \oplus S_2$ is
$S_1^{*} \tilde\oplus S_2^{*}$ and so the operators
$H$
are s.a. restrictions of $S_1^* \tilde\oplus S_2^*$ of the kind
$H_1 \tilde\oplus H_2$.
An interesting point is the relation between the s.a.
extensions of $S_1$, $S_2$ and
the s.a. extensions of
$S_1 \tilde\oplus S_2$.
To characterize the s.a. extensions of $S_1 \tilde
\oplus S_2$ we need the deficiency subspaces
$N_\pm^{1,2}$
of $S_1
\oplus S_2$. Since
\begin{align*}
{N}_{\pm }^{1,2} =& \left
\{ \psi_1\oplus\psi_2 \in D(S_1^*) \oplus D(S_2^*)\,:\,
S_1^* \oplus S_2^* \,(\psi_1\oplus\psi_2)  =
\pm i \,\psi_1\oplus\psi_2  \right\} \\
=& {N}_{\pm }^1 \oplus {N}_{\pm }^2
\end{align*}
where ${N}_{\pm }^1$ and ${N}_{\pm }^2$ are
the deficiency subspaces of $S_1$ and
$S_2$ respectively. If $d=1$ then the deficiency subspaces are finite dimensional and the deficiency
indices satisfy
\begin{equation*}
m_{\pm}=\mbox{dim}\, {N}_{\pm}^{1,2} = \mbox{dim}\,
{N}_{\pm}^{1} + \mbox{dim}\,
{N}_{\pm }^{2}=m_{\pm}^1+m_{\pm}^2\,.
\end{equation*}
We immediately notice that $S_1$ and $S_2$ may have no s.a. extensions (because $m^1_- \not=m_+^1$ and $m^2_- \not=m_+^2$) and yet $S_1 \oplus
S_2$ may have s.a. extensions (which will not be of the form ${H}_1 \oplus {H}_2$). In the case $d>1$ the situation is similar. Here the
deficiency subspaces $N_\pm^1$ and $N_\pm^2$ are infinite dimensional and $S_1$ and $S_2$ have non-self-adjoint maximal extension (see e.g.
\cite{[BEH]}, Theorem 4.7.9) which however could produce self-adjoint extensions of $S_1\oplus S_2$. \par We already know from Remark 2.2 that
if $ H_1$ and $ H_2$ are s.a. extensions of $S_1$ and $S_2$ then $ H_1 \oplus  H_2$ is a s.a. extension of $S_1 \tilde \oplus S_2$. However, the
converse result is not valid, i.e. not all s.a. extensions of $S_1 \oplus S_2$ are of the form $ H_1 \oplus  H_2$ with $ H_1$ and $ H_2$
self-adjoint. Indeed according to von Neumann's theorem \cite{Neumann,Akhiezer}, the s.a. extensions of $S_1 \tilde\oplus S_2 $ are parametrized
by the unitary operators $U:N_+^{1,2}\to N_-^{1,2}$ while the ones of $S_k$ are parametrized by the unitary operators $U_k:N_+^k\to N_-^k$.
Since there are lots of unitary operators $U:N_+^1\oplus N_+^2\to N_-^1\oplus N_-^2$ which are not of the form $U_1\oplus U_2$, there are lots
of extensions of $S_1\oplus S_2$ which are not of the form $H_1\oplus H_2$. It follows from Theorem 2.3 that these are not confining. We shall
designate them by {\it transversal}, because they are associated with boundary conditions relating the wave functions of the two domains. \par

\end{remark}

\end{section}

\begin{section}{Boundary potentials}
Let us now suppose that $H_0$ is given by the Schr\"odinger operator
$H_0=-\Delta+V$. We take $V\in L^\infty(\RE^d)$ so that it induces a
bounded multiplication operator and
$H_0$ is self-adjoint with domain $D(H_0)=H^2(\RE^d)$.
Here $H^2(\RE^d)$, the self-adjointness domain of $-\Delta$,
denotes the Sobolev-Hilbert space, with scalar product
$$
\langle\phi,\varphi\rangle_{H^2(\RE^d)}:=
\langle\Delta\phi,\Delta\varphi\rangle_{L^2(\RE^d)}+
\langle\phi,\varphi\rangle_{L^2(\RE^d)}\,,
$$  of square-integrable functions with square
integrable distributional Laplacian. Thus $H^2(\RE^d)$ coincides with the maximal domain of definition of $H_0$. Then we suppose that
$\Omega\subset\RE^d$ is an open bounded set with a boundary $\Gamma$ which is a smooth embedded $(d-1)$-dimensional manifold.
\begin{remark}
Both our hypotheses on the potential $V$ and the boundary $\Gamma$
could be weakened. We take here the simplest ones in order to avoid too
many technicalities. Regarding the potential we could require
that it is relatively $-\Delta$-bounded with bound $<1$, so that, by
the Kato-Rellich theorem, $H_0$
is still self-adjoint with domain $H^2(\RE^d)$. Regarding the boundary
$\Gamma$ everything continues to hold (with the same proofs) with
$\Gamma$ of class $\C^{1,1}$, i.e. $\Gamma$
is locally the graph of a
$\C^1$ function with Lipschitz derivatives. With some more work
$\Omega$ could be supposed to have no more than a Lipschitz boundary, the minimal requirement in order to define (almost everywhere on $\Gamma$) the normal at the boundary.

\end{remark}
As in the previous section we pose
$$
\Omega_1:=\Omega\,,\qquad \Omega_2:=(\bar\Omega_1)^c
$$
and we denote by $H^n(\Omega_k)$, $k=1,2$, $n$ a positive integer, the Sobolev-Hilbert  space given
by completing the pre-Hilbert space $R_k(\C_c^{\infty}(\RE^d))$
endowed with the scalar product
$$
\langle\phi,\varphi\rangle_{H^n(\Omega_k)}=\sum_{0\le \alpha_1+\dots+\alpha_d\le n} \langle
\partial_{x_1}^{\alpha_1}\dots \partial_{x_d}^{\alpha_d} \phi,\partial_{x_1}^{\alpha_1}\dots \partial_{x_d}^{\alpha_d} \varphi\rangle_{L^2(\Omega_k)}\,.
$$
Analogously $H^n_0(\Omega_k)\subsetneq H^n(\Omega_k)$ denotes  the completion of pre-Hilbert space $\C_c^{\infty}(\Omega_k)$ endowed with the same scalar product as $R_k(\C_c^{\infty}(\RE^d))$.\par
Obviously the extension operator
$$
E_k:\C^\infty_c(\Omega_k)\to \C^\infty_c(\RE^d)
$$
is continuous with respect to the $H^2$-type norms. Since
$\C^\infty_c(\Omega_k)$ and $\C^\infty_c(\RE^d)$ are dense in
$H^2_0(\Omega_k)$ and $H^2(\RE^d)$ respectivley,
$$
E_k:H^2_0(\Omega_k)\to H^2(\RE^d)
$$
and
$$
S_k:H^2_0(\Omega_k)\subset L^2(\Omega_k)\to L^2(\Omega_k)\,, \quad
S_k=R_kH_0E_k=-\Delta_{\Omega_k}+V_k
$$
is a well-defined densely defined closed symmetric operator.
Here
$\Delta_{\Omega_k}$ denotes the distributional Laplacian on
$L^2(\Omega_k)$ and $V_k:=R_kVE_k$.
Moreover $S_k$ has self-adjoint extensions and all such
extensions
can be explicitly characterized in terms of (eventually non-local)
boundary conditions (see \cite{[G],Posilicano2,[PR],[G2]}
and references
therein, the study of boundary value
  problems by means of self-adjoint extensions goes back to
  \cite{[C]} and was further developed in \cite{[V]}; for other recent
results see \cite{[BL]}, \cite{[Ry]}, \cite{[BMNW]}, \cite{[GM1]},
\cite{[GM2]}, \cite{[BGW]}). \par
Thus, since $H_0=-\Delta+V$
satisfies (\ref{commutator}), according to Theorem 2.3
any $H$ of the kind
$H=H_1\tilde\oplus H_2$, where $H_k$ is a self-adjoint extension of
$S_k$, satisfies properties (1) to (3). \par
We want now to recast the above operator $H$ in the form
$$
H=-\Delta+V+B\,,
$$
where $B$ is some singular boundary potential supported on
$\Gamma$. This is always possible. Indeed
$H$ is a
self-adjoint extension of the symmetric
operator $S_1\tilde\oplus S_2$ given by
restricting the
self-adjoint operator
$H_0$ to the dense domain
$H_0^2(\Omega_1)\tilde\oplus H_0^2(\Omega_2)$.
By \cite{[Po]} any self-adjoint extension of a symmetric
restriction of $H_0$ admits the additive representation
$H=-\Delta+V+ B$, where $B$ is some singular boundary operator.
Such operator $B$ could be obtained by using the general theory
developed in \cite{[Po]}, but for the case here
considered we prefer to present a more explicit (although
equivalent) construction. \par
We know that the self-adjoint $H$ is the restriction of
$S_1^*\tilde\oplus
S_2^*$ to $D(H_1)\tilde \oplus D(H_2)$, where $S_k^*$ is
explicitly given by
(see e.g. \cite{[G],Posilicano2,[PR]})
$$
S_k^*:D(\Delta_{\Omega_k}^{\max})\subseteq L^2(\Omega_k)\to L^2(\Omega_k)
\,,\quad S_k^*\psi_k:=-\Delta_{\Omega_k}\psi_k+V_k\psi_k\,,
$$
$$
D(\Delta_{\Omega_k}^{\max}):=\{\psi_k\in L^2(\Omega_k)\,:\,
\Delta_{\Omega_k}\psi_k\in L^2(\Omega_k)\}\,.
$$
It is known that $D(\Delta_{\Omega_k}^{\max})=H^2(\Omega_k)$ if $d=1$,
otherwise $H^2(\Omega_k)$ is strictly contained in
$D(\Delta_{\Omega_k}^{\max})$.
\par
From now on we will use the notation $\D'(M)$ for the space of
distributions on the set $M$ with corresponding test function space
$\D(M)\equiv \C^\infty_c(M)$; $\langle\cdot,\cdot\rangle$ will denote the $\D'(M)$-$\D(M)$ pairing.\par
Let $$\psi=\chi_{\Omega_1}\phi_1+
\chi_{\Omega_2}\phi_2\equiv J(R_1\phi_1\oplus R_2\phi_2)\,,\quad\phi_k\in \D(\RE^d)\,.$$ Then $\psi\in
D(\Delta^\max_{\Omega_1})\tilde\oplus D(\Delta^\max_{\Omega_2})$
and
$$
S_1^*\tilde\oplus S_2^*\psi=-\chi_{\Omega_1}\Delta\phi_1
-\chi_{\Omega_2}\Delta\phi_2+V\psi\,.
$$
By the distributional Leibniz rule,
$$
\Delta(\chi_{\Omega_k}\phi)=
\phi\Delta\chi_{\Omega_k}+2\nabla\chi_{\Omega_k}\cd\nabla\phi
+\chi_{\Omega_k}\Delta\phi\,.
$$
By the Gauss-Green formula one has, for any test function
$\varphi\in\D(\RE^d)$,
$$
\langle\nabla\chi_{\Omega_k},\varphi\rangle=
-\int_{\Omega_k}\nabla\varphi(x)\,dx=-
\int_{\Gamma}n_k(x)\varphi(x)\,d\sigma_{\Gamma}(x)\,,
$$
where $n_k$ denotes the outer (w.r.t. $\Omega_k$) normal at $\Gamma$ and $\sigma_\Gamma$ is the surface measure of $\Gamma$.
Given the continuous linear map
$$\rho:\D(\RE^d)\to \D(\Gamma)\,, \quad
[\rho\varphi](x):=\varphi(x)\,,\quad x\in \Gamma\,,$$
we
define $\delta_\Gamma\in\D'(\RE^d)$ by
$$
\delta_\Gamma:\D(\RE^d)\to\CO\,,\quad
\langle\delta_\Gamma,\varphi\rangle
:=\langle 1,\rho\varphi\rangle\equiv
\int_\Gamma\varphi(x)\,d\sigma_\Gamma(x)\,,
$$
and
$f\delta_{\Gamma}\in \D'(\RE^d)$, $f\in \D'(\Gamma)$,
by
$$
f\delta_\Gamma:\D(\RE^d)\to\CO\,,\quad \langle f\delta_\Gamma,
\varphi\rangle:=\langle f,\rho\varphi\rangle\,.
$$
Then
$$
\nabla\chi_{\Omega_k}=-n_k\,\delta_\Gamma\,,\quad
\Delta\chi_{\Omega_k}=\nabla\cd\nabla\chi_{\Omega_k}=
-\nabla\cd (n_k\,\delta_\Gamma)\,.
$$
and so, for any $\psi=\chi_{\Omega_1}\phi_1+
\chi_{\Omega_2}\phi_2$, $\phi_k\in\D(\RE^d)$, since
$$n\equiv n_1=-n_2\,,$$ one has
\begin{align*}
&S_1^*\tilde\oplus S_2^*\psi=-\Delta\psi+V\psi+(\Delta\psi-\chi_{\Omega_1}\Delta\phi_1
-\chi_{\Omega_2}\Delta\phi_2)\\
=&-\Delta\psi+V\psi-2(\nabla(\phi_1-\phi_2))\cd n\,
\delta_\Gamma-(\phi_1-\phi_2)\,\nabla (n\cd\delta_{\Gamma})\,.
\end{align*}
Now we introduce
$$\gamma_{\Omega_k}^0
:H^2(\Omega_k)\to L^{2}(\Gamma)\,,\quad \gamma_{\Omega_k}^1
:H^2(\Omega_k)\to L^{2}(\Gamma)
$$
defined as the unique continuous linear maps such that,
when $\psi_k=R_k\phi_k$, $\phi_k\in \D(\RE^d)$,
$$
[\gamma^0_{\Omega_k} \psi_k](x)=\phi_k(x)\,,\quad
[\gamma^1_{\Omega_k} \psi_k](x)=n_k(x)\cd\nabla\phi_k(x)\,,
\quad x\in \Gamma\,.
$$
More precisely these trace operators
have range respectively given
by the fractional Sobolev spaces $H^{3/2}(\Gamma)$
and $H^{1/2}(\Gamma)$ (see e.g. \cite{[LM]}, Chapter 1, Section 8.2).
These maps have unique
continuous extensions (see \cite{[LM-AIF]}, Sections 2 and 3,
\cite{[LM]}, Chapter 2, Section 6.5)
$$\hat \gamma_{\Omega_k}^0
: D(\Delta_{\Omega_k}^{\max})\to \D'(\Gamma)\,,\quad \hat \gamma_{\Omega_k}^1
:D(\Delta_{\Omega_k}^{\max})\to \D'(\Gamma) \,.
$$
More precisely these maps have range respectively given
by the dual Sobolev spaces
$H^{-1/2}(\Gamma)$ and $H^{-3/2}(\Gamma)$.
Let us remark that the results we need here and below
from the quoted references \cite{[LM-AIF],[LM]} were proved there in the case of a
bounded set, an hypothesis which $\Omega^c$ does not satisfy;
however what really counts is not the boundedness of
$\Omega^c$
but the compactness of its boundary: one can check that
this alternative hypothesis suffices (see e.g. \cite{[G2]}).\par
One has
$$
H^2_0(\Omega_k)=\{\psi_k\in H^2(\Omega_k)\,:\, \gamma_k^0\psi_k=\gamma_k^1\psi_k=0\}
$$
and, by elliptic
regularity (see \cite{[LM]}, Chapter 2, Section 7.3),
\begin{align}\label{ellreg}
&H^2(\Omega_k)=\{\psi_k\in D(\Delta_{\Omega_k}^{\max})\,:\,
\hat \gamma_{\Omega_k}^0\psi_k\in H^{3/2}(\Gamma)\}\\
=&
\{\psi_k\in D(\Delta_{\Omega_k}^{\max})\,:\, \hat \gamma_{\Omega_k}^1\psi_k-
f_k\hat \gamma_{\Omega_k}^0\psi_k\in H^{1/2}(\Gamma)\}\,,
\end{align}
where $f_k\in \C^\infty(\Gamma)$.\par
By $\hat \gamma_{\Omega_k}^0$ and $\hat \gamma_{\Omega_k}^1$ we can
then define the continuous linear maps
$$
j_\Gamma^0:D(\Delta_{\Omega_1}^{\max})\tilde\oplus D(\Delta_{\Omega_2}^{\max})\to \D'(\Gamma)
\,,\quad j_\Gamma^0\,\psi
:=\hat\gamma_{\Omega_1}^0\psi_1-\hat\gamma_{\Omega_2}^0\psi_2\,,
$$
$$
j_\Gamma^1:D(\Delta_{\Omega_1}^{\max})\tilde\oplus D(\Delta_{\Omega_2}^{\max})\to \D'(\Gamma)
\,,\quad j_\Gamma^1\,\psi
:=\hat\gamma_{\Omega_1}^1\psi_1+\hat\gamma_{\Omega_2}^1\psi_2\,,
$$
which measure the jumps of $\psi=E_1\psi_1+E_2\psi_2$
and of its normal derivative across $\Gamma$. Thus,
for any $\psi=\chi_{\Omega_1}\phi_1+
\chi_{\Omega_2}\phi_2$, $\phi_k\in\D(\RE^d)$, since
\begin{align*}
&\langle(\nabla(\phi_1-\phi_2))\cd n\,
\delta_\Gamma,\varphi\rangle=\langle n\cd\delta_\Gamma,
\varphi\nabla(\phi_1-\phi_2)\rangle\\
=&\int_\Gamma \varphi(x)\,n(x)\cd\nabla(\phi_1-
\phi_2)(x)\,d\sigma_\Gamma(x)=
\langle j_\Gamma^1\psi\,\delta_\Gamma,\varphi\rangle
\end{align*}
and
\begin{align*}
&\langle(\phi_1-\phi_2)\nabla\cd (n\,
\delta_\Gamma),\varphi\rangle=\langle\nabla\cd (n\,
\delta_\Gamma),
\varphi(\phi_1-\phi_2)\rangle\\
=&-\int_\Gamma n(x)\cd\nabla(\varphi(\phi_1-
\phi_2))(x)\,d\sigma_\Gamma(x)\\
=&-\int_\Gamma (\phi_1-
\phi_2)(x)\,n(x)\cd\nabla\varphi(x)\,d\sigma_\Gamma(x)\\
&-
\int_\Gamma n(x)\cd\nabla(\phi_1-
\phi_2)(x)\,\varphi(x)\,d\sigma_\Gamma(x)
\\
=&\langle\nabla\cd(j_\Gamma^0\psi\,n\,\delta_\Gamma),
\varphi\rangle
-\langle j_\Gamma^1\psi\,\delta_\Gamma,\varphi\rangle\,,
\end{align*}
one has
$$
S_1^*\tilde\oplus S_2^*\psi=-\Delta\psi +V -j^1_\Gamma\psi\, \delta_\Gamma-\nabla\cd(j_\Gamma^0\psi\,n\,\delta_{\Gamma})\,.
$$
Since $S_1^*\tilde\oplus S_2^*$ is a closed operator and
$$\D_{k}(\RE^d):=\{\psi_k=R_{k}\phi_k \,,\ \phi_k\in \D(\RE^d)\}$$
is dense, w.r.t. the graph norm, in
$D(S_k^*)\equiv D(\Delta_{\Omega_k}^{\max})$
(see \cite{[LM-AIF]}, Lemma 2.2, \cite{[LM]}, Chapter 2, Section 6.4), the above additive
representation of $S_1^*\tilde\oplus S_2^*$ extends from
$\D_{1}(\RE^d)\tilde\oplus \D_{2}(\RE^d)$ to
$D(\Delta_{\Omega_1}^{\max})\tilde
\oplus D(\Delta_{\Omega_2}^{\max})$ and so
one has the following
\begin{theorem}
$$
S_1^*\tilde\oplus S_2^*:D(\Delta_{\Omega_1}^{\max})\tilde\oplus
D(\Delta_{\Omega_2}^{\max})\subseteq L^2(\RE^d)\to
L^2(\RE^d)\,,
$$
$$
S_1^*\tilde\oplus S_2^*\psi=-\Delta\psi+V\psi-j^1_\Gamma\psi\,
\delta_\Gamma-\nabla\cd(j_\Gamma^0\psi\,n\,\delta_{\Gamma})\,.
$$
\end{theorem}
We can now restrict $S_1^*\tilde\oplus S_2^*$
to $D(H_1)\tilde\oplus D(H_2)$ thus
obtaining a self-adjoint operator $H$ which satisfies conditions (1)
to (3).
In particular we can restrict $S_1^*\tilde\oplus S_2^*$ to the domains
corresponding to the most common local boundary conditions at the
boundary of $\Omega$: Dirichlet, Neumann and Robin boundary conditions i.e. to
$$
D(\Delta_{\Omega_1}^D)\tilde\oplus D(\Delta_{\Omega_2}^D)\,,\quad
D(\Delta_{\Omega_1}^N)\tilde\oplus D(\Delta_{\Omega_2}^N)\quad
\text{\rm and}\quad
D(\Delta_{\Omega_1}^R)\tilde\oplus D(\Delta_{\Omega_2}^R)\,,
$$
 where
$$
D(\Delta_{\Omega_k}^D):=\{\psi_k\in H^2(\Omega_k)\,:\,
\gamma_{\Omega_k}^0\psi_k=0\}\,,
$$
$$
D(\Delta_{\Omega_k}^N):=\{\psi_k\in H^2(\Omega_k)\,:\,
\gamma_{\Omega_k}^1\psi_k=0\}\,,
$$
$$
D(\Delta_{\Omega_k}^R):=\{\psi_k\in H^2(\Omega_k)\,:\,
\gamma_{\Omega_k}^1\psi_k=f_k\gamma_{\Omega_k}^0\psi_k\}\,,
$$
$f_k\in \C^\infty(\Gamma)$, $f_k=f_k^*$.
We denote by $H^{D}$, $H^N$
and
$H^{R}$ the corresponding self-adjoint operators. \par
Now we use the representation of $S^*_1\tilde\oplus S^*_2$
obtained above to recast any separating self-adjoint
extension  $$
H_1\tilde\oplus H_2
$$
in the form
$$
H=-\Delta+V+B\,,
$$
with $H$ defined on its maximal domain.\par
Let
$$
F_k:D(\Delta_{\Omega_k}^\max)\to \D'(\Gamma)
$$
be any map such that
$$
D(H_k)={\rm Ker}(F_k)\,.
$$
Then
\begin{theorem}
$H_1\tilde\oplus H_2=H$, where
$$
H:D(H)\subseteq L^2(\RE^d)\to L^2(\RE^d)\,,\quad H=-\Delta+V+B\,,
$$
$$
D(H)=\{\psi\in D(\Delta_{\Omega_1}^\max)\tilde \oplus D(\Delta_{\Omega_2}^\max)\,:\,-\Delta\psi+V\psi+B\psi\in L^2(\RE^d)\}\,,
$$
$$
B\psi=F\psi-j^1_\Gamma\psi\,
\delta_\Gamma-\nabla\cd(j_\Gamma^0\psi\,n\,\delta_{\Gamma})
$$
$$
F\psi:=c_1(F_1\psi_1+F_2\psi_2)\delta_\Gamma+c_2\nabla\cd((F_1\psi_1-F_2\psi_2)n\delta_\Gamma)\,,\quad
c_1c_2\not=0\,.
$$
\end{theorem}
\begin{proof} For any $\psi=\psi_1\tilde\oplus\psi_2\in
D(\Delta_{\Omega_1}^\max)\tilde \oplus D(\Delta_{\Omega_2}^\max)$ one has
$$
-\Delta\psi+V\psi+B\psi=S_1^*\tilde\oplus S_2^*\psi+F\psi
$$
and so
\begin{align*}
&-\Delta\psi+V\psi+B\psi\in L^2(\RE^d)\ \iff\ F\psi=0\\ &\iff \psi_k\in {\rm Ker}(F_k)\,,\ k=1,2\,,
\iff \psi \in D(H_1\tilde\oplus H_2)\,,
\end{align*}
i.e. $D(H)=D(H_1\tilde\oplus H_2)$. The proof is then concluded by
$H\subseteq S_1^*\tilde\oplus S_2^*$.
\end{proof}
Let us define the boundary singular potential
$$B^{D}:D(\Delta_{\Omega_1}^{\max})\tilde
\oplus D(\Delta_{\Omega_2}^{\max})\to\D'(\RE^d)\,,
\quad B^{D}\psi:=(\mu^0_\Gamma\psi\,-j^1_\Gamma\psi)\,
\delta_\Gamma\,,
$$
where
$$
\mu^0_\Gamma:D(\Delta_{\Omega_1}^{\max})\tilde
\oplus D(\Delta_{\Omega_2}^{\max})\to \D'(\Gamma)\,,
\quad\mu^0_\Gamma\psi:=
\frac{1}{2}\,\left(\hat\gamma_{\Omega_1}^0\psi_1+
\hat\gamma_{\Omega_2}^0\psi_2\right)\,,
$$
is the mean of the inner and outer limits at $\Gamma$ of
$\psi=E_1\psi_1+E_2\psi_2$. Then
\begin{corollary}
$$
H^{D}:D(H^{D})\subseteq L^2(\RE^d)\to L^2(\RE^d)\,,\quad
H^{D}=-\Delta+V+B^{D}\,,
$$
$$
D(H^{D})=\{\psi\in D(\Delta_{\Omega_1}^{\max})
\tilde\oplus D(\Delta_{\Omega_2}^{\max})\,:\,
-\Delta\psi+V\psi+B^{D}\psi\in L^2(\RE^d)\}\,.
$$
\end{corollary}
\begin{proof}
By taking $F_k=F_k^D$,
$$
F_k^D:D(\Delta_{\Omega_k}^\max)\to \D'(\Gamma)\,,
\quad F^D_k\psi_k:=\hat\gamma^0_{\Omega_k}\psi_k\,,
$$
one has
$$
D(\Delta_{\Omega_k}^D)={\rm Ker}(F^D_k)
$$
by elliptic regularity. The proof is then concluded by
\begin{align*}
&\frac{1}{2}\,(F^D_1\psi_1+F^D_2\psi_2)
\delta_\Gamma+\nabla\cd((F^D_1\psi_1-F^D_2\psi_2)n\delta_\Gamma)\\
&-j^1_\Gamma\psi\,
\delta_\Gamma-\nabla\cd(j_\Gamma^0\psi\,n\,\delta_{\Gamma})\\
=&B^D\psi\,.
\end{align*}
\end{proof}
A similar result can be obtained in the case of Neumann
boundary conditions. In this case we
define the boundary singular potential
$$
B^{N}:D(\Delta_{\Omega_1}^{\max})\tilde
\oplus D(\Delta_{\Omega_2}^{\max})\to\D'(\RE^d)\,,
\quad B^{N}\psi:=\nabla\cd((\mu^1_\Gamma\psi\,-j^0_\Gamma\psi)\,
n\delta_\Gamma)\,,
$$
where
$$
\mu^1_\Gamma:D(\Delta_{\Omega_1}^{\max})\tilde
\oplus D(\Delta_{\Omega_2}^{\max})\to \D'(\Gamma)\,,
\quad\mu^1_\Gamma\psi:=
\frac{1}{2}\,\left(\hat\gamma_{\Omega_1}^1\psi_1-
\hat\gamma_{\Omega_2}^1\psi_2\right)\,,
$$
is the mean of the inner and outer limits at $\Gamma$ of the normal derivative of $\psi=E_1\psi_1+E_2\psi_2$.
Then
\begin{corollary}
$$
H^{N}:D(H^{N})\subseteq L^2(\RE^d)\to L^2(\RE^d)\,,\quad
H^{N}=-\Delta+V+B^{N}\,,
$$
$$
D(H^{N})=\{\psi\in D(\Delta_{\Omega_1}^{\max})
\tilde\oplus D(\Delta_{\Omega_2}^{\max})\,:\,
-\Delta\psi+V\psi+B^{N}\psi\in L^2(\RE^d)\}\,.
$$
\end{corollary}
\begin{proof}
By taking $F_k=F_k^N$,
$$
F_k^N:D(\Delta_{\Omega_k}^\max)\to \D'(\Gamma)\,,
\quad F^N_k\psi_k:=\hat\gamma^1_{\Omega_k}\psi_k\,,
$$
one has
$$
D(\Delta_{\Omega_k}^N)={\rm Ker}(F^N_k)
$$
by elliptic regularity. The proof is then concluded by
\begin{align*}
&(F^N_1\psi_1+F^N_2\psi_2)
\delta_\Gamma+\frac{1}{2}\,\nabla\cd((F^N_1\psi_1-F^N_2\psi_2)n\delta_\Gamma)\\
&-j^1_\Gamma\psi\,
\delta_\Gamma-\nabla\cd(j_\Gamma^0\psi\,n\,\delta_{\Gamma})\\
=&B^N\psi\,.
\end{align*}
\end{proof}
The case of Robin boundary conditions is no more than a
combination
of the two preceding ones by taking
$$
F_k=F_k^R:=F_k^N-f_kF_k^D\,.
$$
So, by defining
$$
B^{R}:
D(\Delta_{\Omega_1}^{\max})\tilde\oplus D(\Delta_{\Omega_2}^{\max})\to\D'(\RE^d)\,,
$$
$$
B^{R}\psi:=\nabla\cd((\mu_\Gamma^1\psi
-\mu_\Gamma^{0,f_1,f_2}\psi-j_\Gamma^0\psi)n\,\delta_\Gamma)-
j_\Gamma^{0,f_1,f_2}\psi\,\delta_\Gamma
\,,
$$
where
$$
\mu^{0,f_1,f_2}_\Gamma:D(\Delta_{\Omega_1}^{\max})\tilde\oplus
D(\Delta_{\Omega_2}^{\max})\to\D'(\RE^d)\,,\quad
\mu^{0,f_1,f_2}_\Gamma\psi:=
\frac{1}{2}\,\left(f_1\hat\gamma_{\Omega_1}^0\psi_1-
f_2\hat\gamma_{\Omega_2}^0\psi_2\right)\,,
$$
$$
j^{0,f_1,f_2}_\Gamma:D(\Delta_{\Omega_1}^{\max})\tilde\oplus
D(\Delta_{\Omega_2}^{\max})
\to \D'(\Gamma)\,,\quad
j^{0,f_1,f_2}_\Gamma\psi
:=f_1\hat\gamma_{\Omega_1}^0\psi_1+f_2\hat\gamma_{\Omega_2}^0\psi_2
\,,
$$
one has
\begin{corollary}
$$
H^{R}:D(H^{R})\subset L^2(\RE^d)\to L^2(\RE^d)\,,\quad
H^{R}=-\Delta+V+B^{R}\,,
$$
$$
D(H^{R})=
\{\psi\in D(\Delta_{\Omega_1}^{\max})
\tilde\oplus D(\Delta_{\Omega_2}^{\max})\,:\,
-\Delta\psi+V\psi+B^{R}\psi\in L^2(\RE^d)\}\,.
$$
\end{corollary}
We conclude by discussing the case of a general (non-local)
boundary conditions.
\par
The most general separating self-adjoint extension are of the kind $H_1\tilde\oplus H_2$,
with (see \cite{[G],Posilicano2,[PR],[G2]} and references
therein)
$$
H_k:D(H_k)\subseteq L^2(\Omega_k)\to L^2(\Omega_k)\,,\quad
H_k\psi_k:=-\Delta_{\Omega_k}\psi_k+V_k\psi_k\,,
$$
\begin{align*}
D(H_k)=&\{\psi_k\in D(\Delta_{\Omega_k}^\max)\,:\,
\Sigma\hat\gamma^0_{\Omega_k}\psi_k\in D(\Theta_k)\,,\ \\
&\Pi_k(\hat\gamma_{\Omega_k}^1\psi_k-P^{DN}_{\Omega_k}\hat\gamma^0_{\Omega_k}\psi_k)=\Theta_k\Sigma\hat\gamma^0_{\Omega_k}\psi_k\}\,,
\end{align*}
where $\Pi_k$ is an orthogonal projector in $H^{1/2}(\Gamma)$,
$\Sigma$ is the canonical
unitary isomorphism mapping the
space $H^{s}(\Gamma)$ onto $H^{s+1}(\Gamma)$,
$\Theta_k$
is a self-adjoint operator acting in the Hilbert space given by the
the range of $\Pi_k$,
$P^{DN}_{\Omega_k}:H^{s}(\Gamma)\to H^{s-1}(\Gamma)$ is the
Dirichlet-to-Neumann operator (relative to $\Omega_k$) over
$\Gamma$. Let us remark that for any $\psi_k\in D(\Delta_{\Omega_k}^\max)$ the difference
$\hat\gamma_{\Omega_k}^1\psi_k-P_{\Omega_k}^{DN}\hat\gamma_{\Omega_k}^0\psi_k$ is always
$H^{1/2}(\Gamma)$-valued.\par
Note that Dirichlet boundary conditions
correspond to $\Pi_k=0$ and Robin boundary conditions
correspond to $\Pi_k=1$ and $\Theta_k=\Theta_k^R$,
$$\Theta_k^R:=(-P^{DN}_{\Omega_k}+M_k)\Sigma^{-1}:H^{5/2}(\Gamma)\subset H^{1/2}(\Gamma)\to H^{1/2}(\Gamma)\,,$$
where $M_k$ denotes the multiplication by $f_k$.\par
For general boundary conditions one is lead to take
$F_k=\hat F_k$, where
$$
\hat F_k\psi_k:=\hat\gamma^0_{\Omega_k}\psi_k-
\Sigma^{-1}(\Theta_k+i)^{-1}\Pi_k
(\hat\gamma^1_{\Omega_k}\psi_k-P^{DN}_{\Omega_k}\hat\gamma^0_{\Omega_k}\psi_k-i\Sigma\hat\gamma^0_{\Omega_k}\psi_k)\,.
$$
This gives $\hat F_k=F_k^D$ when $\Pi_k=0$, but produces
$$\hat F_k\psi_k=\Sigma^{-1}(\Theta^R_k+i)^{-1}
(f_k\hat\gamma^0_{\Omega_k}\psi_k-\hat\gamma^1_{\Omega_k}\psi_k)
\not=F_k^R
$$ in the case of Robin boundary conditions.  This shows
that there exist different confining potentials for the same
boundary conditions, some being more convenient than others.
\begin{remark}
Is it possible to use alternative
(i.e. which use different extension parameters)
representations of $D(H_k)$, as suggested by boundary triple
theory (see e.g. \cite{[GG]} and references therein).
This produces similar results.
For example one can write
\begin{align*}
D(H_k)=\{&\psi_k\in D(\Delta_{\Omega_k}^\max)\,:\,
i(1+U_k)\Sigma \hat\gamma^0_{\Omega_k}\psi_k\\
=&(1-U_k)(\hat\gamma_{\Omega_k}^1\psi_k-P^{DN}_{\Omega_k}\hat\gamma^0_{\Omega_k}\psi_k)\}\,,
\end{align*}
where $$U_k:H^{1/2}(\Gamma)\to H^{1/2}(\Gamma)$$ is unitary. In this case
Dirichlet boundary conditions correspond to the choice $U_k=1$ and
Robin ones to $U_k=(\Theta_k^R-i)(\Theta^R_k+i)^{-1}$, i.e. $U_k$ is the Cayley transform of $\Theta_k^R$.
\end{remark}
\end{section}

%%% Applicazioni
\begin{section}{An application: Deformation quantization of 1-dimensional systems with boundaries}

Let
$$H_0:H^2(\RE)\subset L^2(\RE)\to L^2(\RE)\,,\qquad
H_0=-\Delta=\left(-i\,\frac{d\ }{dx}\right)^2$$ be the self-adjoint
Hamiltonian for the free
particle in one dimension. Let us pose
$$\Omega_1=\Omega:=(a,b)\,,\quad \Omega_2=
\Omega_-\cup \Omega_+\,,\quad \Omega_-:=(-\infty,a)\,,\quad
\Omega_+:=(b,+\infty)\,.$$ Our first aim is to determine the
explicit form of the operator $H^D$, associated to $H_0$,
describing the system confined to $\Omega$ and satisfying
Dirichlet boundary conditions at $a$ and $b$. Since for any open
${\mathcal O}\subseteq\RE$ $D(\Delta_{{\mathcal
O}}^{max})=H^2({\mathcal O})$, one has
$$
D(\Delta_{\Omega_1}^{max})\tilde\oplus D(\Delta_{\Omega_2}^{max})=
H^2(\Omega_-)\tilde\oplus H^2(\Omega)
\tilde\oplus H^2(\Omega_+)=H^2_{a,b}(\RE)\,,
$$
where
$$
H^2_{a,b}(\RE):=\{
\psi\in L^2(\RE)\,:\,\psi=\sum_{{\mathcal O}=\Omega_-,\Omega,\Omega_+}
\chi_{{\mathcal O}}\,\phi_{\mathcal O}\,,\
\phi_{\mathcal O}\in H^2(\RE)\}\,.
$$
Moreover, since
$H^1(\RE)\subset \C_b(\RE)$, the left and right limits
exist and are finite; so for any $\psi\in H^2_{a,b}(\RE)$ we can define the mean $\langle \psi\rangle_y$ and the jump $[\psi']_y$
at $y=a,b$ by
$$
\langle \psi\rangle_y:=\frac{1}{2}\,
\left(\lim_{x\to y-}\psi(x)+\lim_{x\to y+}\psi(x)\right)\,,
$$
$$
[\psi']_y:=\lim_{x\to y+}\psi'(x)-
\lim_{x\to y-}\psi'(x)\,.
$$
Thus, by the results provided in Section 3 applied to this simple 1-dimensional case,
\begin{align*}
&(-\Delta^{max}_{\Omega_1})\tilde\oplus(-\Delta^{max}_{\Omega_2})\psi
=(-\Delta^{max}_{\Omega_-})\tilde\oplus(-\Delta^{max}_\Omega)\tilde\oplus(-\Delta^{max}_{\Omega_+})\psi\\
&=-\psi''+[\psi']_a\,\delta_a+[\psi']_b\,\delta_b+
[\psi]_a\,\delta_a'+[\psi]_b\,\delta_b'\,,
\end{align*}
and
$$
(-\Delta_{\Omega_1}^D)\tilde\oplus(-\Delta_{\Omega_2}^D)=
(-\Delta_{\Omega_-}^D)\tilde\oplus(-\Delta_{\Omega}^D)
\tilde\oplus(-\Delta_{\Omega_+}^D)=
H^D\,,
$$
where $H^D$ has the two equivalent representations
$$
H^D\psi=-\psi''+B^D_0\psi\,,\quad B^D_0\psi=[\psi']_a\,\delta_a+[\psi']_b\,\delta_b\,,
$$
\begin{align*}
D(H^D)=\{\psi\in H^2_{a,b}(\RE)\cap \C(\RE)\,:\,\psi(a)=\psi(b)=0\}\,,
\end{align*}
and
$$
H^D\psi=-\psi''+B^D\psi\,,\quad B^D\psi:=
\left(\langle\psi\rangle_a+[\psi']_a\right)\,\delta_a+
\left(\langle\psi\rangle_b+[\psi']_b\right)\,\delta_b\,,
$$
\begin{align*}
D(H^D)=\{\psi\in H^2_{a,b}(\RE)
\,:\, H^D\psi\in L^2(\RE)\}\,.
\end{align*}
Since
$$
\sigma(-\Delta_{\Omega_\pm}^D)=\sigma_{ac}(-\Delta_{\Omega_\pm}^D)=
[0,\infty)\,, $$
$$
\sigma(-\Delta_{\Omega}^D)=\sigma_{disc}(-\Delta_{\Omega}^D)=
\{\lambda_n^2\}_1^\infty\,,\quad\lambda_n=
\frac{n\pi}{b-a}\,,
$$
one has
$$
\sigma_{disc}(H^D)=\emptyset\,,\quad
\sigma_{point}(H^D)=\{\lambda^2_n\}_1^\infty\,,\quad
\sigma_{ac}(H^D)=[0,\infty)\,.
$$
The eigenfunction of $H^D$ corresponding to the
eigenvalue $\lambda_n^2$ is $$\psi_n=J(0\oplus\phi_n\oplus 0)\,,$$ where $\phi_n$ is the eigenfunction of $-\Delta^D_\Omega$ corresponding to $\lambda_n^2$, i.e.
\begin{equation}\label{eigenfunction}
\psi_n(x)=
\chi_{(a,b)}(x)\sin \lambda_n(x-a)\,.
\end{equation}
Notice that the same eigenfunctions $\psi_n$ could have been obtained by at first solving the unconstrained equation $H_0\psi=\lambda\psi$,
then by imposing the boundary conditions $\psi(a)=\psi(b)=0$ and finally by using the projection $P_\Omega$ i.e. by taking the product with $\chi_{(a,b)}$. As we will see below this equivalence between the two methods for obtaining the eigenfunctions of the constrained systems fails in the deformation quantization framework. Only the global (i.e. in terms of $H^D$) formulation of the
eigenvalue problem provides the correct results. This is an
interesting example of the important role played by the boundary
potentials in the context of non-local formulations of quantum
mechanics (like the deformation quantization).\par
Deformation quantization provides an alternative, autonomous
formulation of quantum mechanics in terms of phase space objects.
The deformation approach is valid for a large class of systems,
including those defined on (non-flat) sympletic \cite{Fedosov} and
Poisson \cite{Kontsevich} manifolds. In the simplest flat case the
theory yields the Weyl-Wigner formulation of quantum mechanics
\cite{Bayen}. In this context, the eigenvalue problem (for a
generic observable $a$) is formulated in terms of the left
and right stargenvalue (or $*$-genvalue) equations
\begin{equation}\label{*eigenvalue}
a*f^W_{\alpha}=f^W_\alpha*a=\alpha
f^W_\alpha
\end{equation}
where $a:\RE^2\to\RE$ is a real phase space function (or distribution), $f^W_\alpha:\RE^2\to\RE$ is the stargenfunction associated to the eigenvalue $\alpha$
and $*$ is the star product. The latter is the fundamental algebraic structure of the quantum phase space, and is most naturally defined in
terms of the Weyl-Wigner transform $W$. \par
Let us denote by $\L(\S(\RE^n),\S'(\RE^n))$ the space of continuous linear maps on $\S(\RE^n)$, the space of Schwartz test functions, to $\S'(\RE^n)$, the space of tempered distributions.
Then
$$
W:\L(\S(\RE),\S'(\RE))\to \S'(\RE^2)
$$
is the continuous linear one-to-one invertible map which associates to $A\in\L(\S(\RE),\S'(\RE))$ its symbol $a\equiv W(A)\in \S'(\RE^2)$ defined by
\begin{equation}\label{WW}
\forall \phi\,,\,\varphi\in\S(\RE)\,,\quad
\langle W(A),\phi\otimes\varphi\rangle=
\langle A_K,\phi\otimes\hat\varphi\circ T\rangle\,.
\end{equation}
Here $\hat\varphi(y)=\smallint_{-\infty}^{+\infty} e^{-iyp}\varphi(p)\,dp$ denotes Fourier transform,
$$T(x,y):=\left(\frac{x+y}{2},x-y\right)\,, $$
and we used the identification (given by the Schwartz's kernel theorem) $\L(\S(\RE),\S'(\RE))\simeq\S'(\RE)\otimes\S'(\RE)\simeq \S'(\RE^2)$ which associates to any $A$ its kernel $A_K$.
In the case the kernel $A_K$ is in $\S(\RE^2)$ one has (see e.g. \cite{Bracken,Dubin,Wong})
$$
W(A)(x,p)
:=  \int_{-\infty}^{+\infty} e^{-ipy}A_K\left(x+\frac{y}{2},x-\frac{y}{2}\right)\,dy\,.
$$
The star product of two symbols $a,\,b\in\S'(\RE^2)$ such that $AB\in \L(\S(\RE),\S'(\RE))$, $A:=W^{-1}(a)$, $B:=W^{-1}(b)$, is then defined by
\begin{equation}\label{*1}
a*b:=W(AB)\,.
\end{equation}
In the case $a,\,b\in\S(\RE^2)$ one has the more explicit
formula (Moyal product)
$$
a*b\,(z)=\left(\frac{1}{4\pi}\right)^2\int_{\RE^2\times \RE^2}e^{-i\omega(u_1,u_2)/2}\,a\left(z+\frac {u_1}2\right)b\left(z-\frac {u_2}2\right)\, du_1du_2\,,
$$
where $z\equiv(x,p)$, $u_k\equiv(q_k,p_k)$, $k=1,2$, and $\omega$
denotes the standard symplectic form
$\omega(u_1,u_2):=q_1p_2-q_2p_1$.\par
We now come back to the
eigenvalue problem (\ref{*eigenvalue}). It was proved for a large
class of Hamiltonians (and particularly for all of the form
$H=-\Delta+V$, where $V\in \C^{\infty}(\RE)$) that if $\psi_E$ is
the general solution of the eigenvalue equation $H \psi_E=E\psi_E$
then the Wigner function associated to $\psi_E$, i.e. the Weyl-Wigner transform of the orthogonal projector $|\psi_E\rangle\langle\psi_E|$
$$
f_E^W(x,p)=\frac{1}{2\pi}W(|\psi_E\rangle\langle\psi_E|)(x,p)=\frac{1}{\pi} \int_{-\infty}^{+\infty}
e^{-2ipy} \psi^*_E(x-y)\psi_E(x+y)\,dy
$$
is a solution of the $*$-genvalue equations
$$
h*f_E^W=f_E^W*h=Ef_E^W
$$
for $h=W(H)$ \cite{Fairlie1,Bayen,Dubin,Dias5,Gosson}. The
converse result, showing that all energy stargenfunctions for all
Hamiltonians of the form $H=-\Delta+V$, with $V\in
\C^{\infty}(\RE)$, are the Wigner functions of the energy
eigenfunctions was developed originally in \cite{Baker,Fairlie1}.
A more rigorous and general derivation can be found in
\cite{Gosson}.
\par
Unfortunately, this elegant relation does not extend to the case
of systems with boundaries. More precisely, if we address the
confined eigenvalue problem, in the deformation context, by using
the standard approach of solving for the unconstrained system and
projecting the solution into the bulk then we are faced with a
surprising problem. In fact, the Wigner function associated to the
confined eigenfunction (which is, by construction, the correct
stargenfunction of the confined system) does not satisfy the
stargenvalue equation for the unconstrained system, not even in
the bulk \cite{Dias1,Kryukov,Walton,Dias6}.\par Let us denote by
$f^W_n$ the Wigner function associated to the state $\psi_n$ (see
(\ref{eigenfunction})) with corresponding eigenvalue
$\lambda_n^2$. Since $W(-\Delta)=p^2$, we have by (\ref{*1}),
$$\frac{1}{2\pi}\,W(-\Delta\,|\psi_n\rangle\langle\psi_n|\,)=p^2*f^W_n\,. $$
Since
$$
H^D\psi_n=\lambda_n^2\psi_n\iff H^D\,|\psi_n\rangle\langle\psi_n|=\lambda_n^2\,
\,|\psi_n\rangle\langle\psi_n|\,\,,
$$
and both $H^D\,|\psi_n\rangle\langle\psi_n|\,=-\Delta\,|\psi_n\rangle\langle\psi_n|\,
+B^D\,|\psi_n\rangle\langle\psi_n|\,$ and $B^D\,|\psi_n\rangle\langle\psi_n|\,\equiv B^D_0\,|\psi_n\rangle\langle\psi_n|\,$ are in
$\L(\S(\RE),\S'(\RE))$, one has
$$
\frac{1}{2\pi}\,W(H^D\,|\psi_n\rangle\langle\psi_n|\,)=
p^2*f^W_n+\frac{1}{2\pi}\,W(B^D_0\,|\psi_n\rangle\langle\psi_n|\,)
=\lambda_n^2 f^W_n\,.
$$
Since
$$
(B^D_0\,|\psi_n\rangle\langle\psi_n|\,)_K=\left(\psi_n'(a+)\delta_a-\psi_n'(b-)
\delta_b\right)\otimes\psi^*_n\,,
$$
by the definition of Weyl-Wigner transform (\ref{WW})
one obtains
\begin{align*}
&\frac{1}{2\pi}W(B^D_0\,|\psi_n\rangle\langle\psi_n|\,)(x,p)\\
=&\frac{1}{\pi} \left(\psi_n'(a+)e^{-2ip(a-x)}\psi^*_n(2x-a)-
\psi_n'(b-)e^{-2ip(b-x)}\psi^*_n(2x-b)\right)\,.
\end{align*}
So, posing $b^D_n:=\frac{1}{2\pi}\,W(B^D\,|\psi_n\rangle\langle\psi_n|\,)$,
one has that any $*$-genfunction $f^W_n$ must solve the equation
\begin{equation}\label{equation}
p^2*f^W_n+b_n^D=\lambda_n^2 f^W_n\,.
\end{equation}
This reproduces the result obtained in \cite{Dias1}, equation
(29).\par Since $b^D_n(x,p)\not=0$ for $a<x<b$, (\ref{equation})
shows that, contrarily to the eigenfunctions $\psi_n$, the
corresponding $*$-genfunctions $f^W_n$ can not satisfy the
equation $p^2*f^W_n=\lambda_n^2 f^W_n$, not even in the bulk.
Thus, as we already remarked, only a global formulation in terms
of $H^D$ can provide the correct results.\par Now we would like to
interpret (\ref{equation}) as a $*$-eingenvalue equation. However
it is not possible to write $b_n^D=b_0^D*f^W_n$ with
$b_0^D=W(B^D_0)$ since $B_0^D$ is not a generalized operator from
$\S(\RE)$ to $\S'(\RE)$ (notice that the restriction to $\S(\RE)$
of $B_0^D$ is the zero operator). To address this problem, we need
to revisit the definition of $B_0^D$.\par Let us define
$B^D_\pm\in\L(\S(\RE),\S'(\RE))$ by
$$
B^D_\pm\psi:=\mp\psi\,(\delta'_a+\delta'_b)\,.
$$
Notice that the action of $B^D_\pm$ can be extended to the set of
all functions $\psi$ which are continuous and differentiable in a
neighborhood of both $a$ and $b$; we have
$$
B^D_\pm\psi=\pm(\psi'(a)\,\delta_a+\psi'(b)\,\delta_b)\mp(\psi(a)\,\delta'_a+\psi(b)\,\delta'_b)\,.
$$
Thus for any $\psi \in D(H^D)$, i.e. for any $\psi\in
H^2_{a,b}(\RE)\cap \C(\RE)$ such that $\psi(a)=\psi(b)=0$, one has
\begin{align*}
B^D_0\psi & =  [\psi' ]_a\,\delta_a+ [\psi']_b\,\delta_b \\
&= \left(\lim_{x\to a^+} \psi'(x)-\lim_{x \to a^-}
\psi'(x) \right)\,\delta_a + \left(\lim_{x\to b^+} \psi'(x)-\lim_{x
\to b^-} \psi'(x) \right)\,\delta_b\\
&=\lim_{\epsilon\downarrow 0}\, (\psi'_{+\epsilon}-\psi'_{-\epsilon})\,(\delta_a
+\delta_b)\\
&= \lim_{\epsilon\downarrow 0}\,
(B^D_+\psi_{+\epsilon}+B^D_-\psi_{-\epsilon}) \,,\quad
\psi_{\pm\epsilon}(x):=\psi(x\pm\epsilon)\,,
\end{align*}
where the limits $\epsilon\downarrow 0$ are taken in $\S'(\RE)$. We now use this formulation of $B^D_0$ to
determine the boundary potential term in the $*$-genvalue
equation. Since, by (\ref{*1}),
\begin{align*}
&W(B^D_+\,|\psi_{n,+\epsilon}\rangle\langle\psi_{n,+\epsilon}|\,+B^D_-
\,|\psi_{n,-\epsilon}\rangle\langle\psi_{n,-\epsilon}|\,)\\
=&
W(B^D_+)*W(\,|\psi_{n,+\epsilon}\rangle\langle\psi_{n,+\epsilon}|\,)
+W(B^D_-)*W(\,|\psi_{n,-\epsilon}\rangle\langle\psi_{n,-\epsilon}|\,)
\end{align*}
and
\begin{align*}
&
\frac{1}{2\pi}\lim_{\epsilon\downarrow 0}\,W(B^D_+\,|\psi_{n,+\epsilon}\rangle\langle\psi_{n,+\epsilon}|\,+B^D_-\,|\psi_{n,-\epsilon}\rangle\langle\psi_{n,-\epsilon}|\,)(x,p)\\
=&\frac{1}{\pi}\lim_{\epsilon\downarrow 0}\, \psi_n'(a+\epsilon)e^{-2ip(a-x)}\psi_n^*(2x-a+\epsilon)\\
&-\frac{1}{\pi}\lim_{\epsilon\downarrow 0}\,
\psi_n'(b-\epsilon)e^{-2ip(b-x)}\psi^*_n(2x-b-\epsilon)
\\
=&b^D_n(x,p)\,,
\end{align*}
posing
$$
b^D_\pm:=W(B^D_\pm)=\mp(\delta'_a+\delta'_b)\,,\quad
f^W_{n,\pm\epsilon}:=\frac{1}{2\pi}\,W
(\,|\psi_{n,\pm\epsilon}\rangle\langle\psi_{n,\pm\epsilon}|\,)\,,
$$
one obtains the $*$-genvalue equation
$$
p^2*f_n^W+\lim_{\epsilon\downarrow 0}\,(b^D_+*f^W_{n,+\epsilon})+
\lim_{\epsilon\downarrow 0}\,(b^D_-*f^W_{n,-\epsilon})=\lambda_n^2f^W_n\,.
$$
Hence, the relation between eigenfunctions and stargenfunctions is
restored for systems with boundaries, provided the global (i.e.
the boundary potential) formulation is used. Equivalent results
can be obtained for the other representation of $H^D$.

\end{section}

\section*{Acknowledgments}
N.C.D. and J.N.P. thank P. Garbaczewski for several discussions. N.C.D. and J.N.P. were partially supported by grant PTDC/MAT/ 69635/2006 of the
Portuguese Science Foundation.

\end{document}